\documentclass[aps,prd,twocolumn,groupedaddress,showkeys,showpacs,superscriptaddress]{revtex4}
\usepackage{graphicx}
\usepackage{natbib}

\newcommand{\feyn}[4]{\hline $#1$ & $#2$ & $#3$ & $#4$ \\}

\begin{document}

\title{Dark matter in a Simplest Little Higgs with T-parity model}

\author{Pat Kalyniak}
\affiliation{Ottawa Carleton Institute for Physics, Department of Physics, Carleton University, 1125 Colonel By Drive, Ottawa K1S5B6 Canada}
\author{Dmitriy Tseliakhovich}
\affiliation{California Institute of Technology, Mail Code 249-17, Pasadena, CA 91125}

\date{\today}

\begin{abstract}
Little Higgs models may provide a viable alternative to supersymmetry as an extension of the Standard Model. After the introduction of a discrete $Z_2$ symmetry, dubbed T-parity into Little Higgs models they also contain a promising dark matter candidate. We investigate a heavy neutrino as a dark matter candidate in the Simplest Little Higgs with T-parity (SLHT) model. First we calculate relic density constraints on the parameter space of the model and then discuss collider signatures for the SLHT dark matter.
\end{abstract}
\pacs{95.35.+d, 12.60.-i}
\keywords{Dark matter, LHC, Simplest Little Higgs}

\maketitle

\section{Introduction}

The combined effort of theoretical and experimental research in physics and astronomy has resulted in the formulation of two extremely successful models that describe most observed phenomena to date. The current knowledge in particle physics is summarized in what is known as the Standard Model~\cite{sm} while the results of the astrophysical observations are described within the Cold Dark Matter with dark energy ($\Lambda$CDM) paradigm. There is, however, broad acceptance that both models are incomplete. One particular problem inherent to both models is the inability to coherently explain the phenomenon of dark matter.  

Dark matter dominates the mass budget of the universe~\cite{Spergel07}, but the actual particle(s) that comprise dark matter are still unknown. Dark matter has yet to be directly detected experimentally, despite attempts to intercept it passing through Earth and attempts to generate it via particle collisions. The broadest constraints on the properties of dark matter result from astronomical observations, which, over the previous decades, have supported what is known as the Cold Dark Matter (CDM) paradigm~\cite{Blumenthal84}. CDM is a phenomenological model, specifying a perfectly ``cold'' (i.e., without initial random motions) particle that interacts only through gravity, and is consequently also dissipationless. There are no astronomical observations or particle physics experiments that rule out a cold, non-interacting dark matter particle, and many observations that support it. However, some theoretical predictions based on CDM models are mildly at odds with observations probing the kpc-scale distribution of dark matter (e.g., \cite{Sellwood02}). More importantly no particle physics model produces a dark matter candidate conforming to the ideal of the CDM paradigm. Specifically, no particle can be perfectly cold or perfectly non-interacting. 

While the Standard Model (SM) of particle physics does not provide a natural non-baryonic dark matter candidate, the near-universal acceptance of dark matter as a major constituent of the universe has inspired many extensions to the Standard Model that, in addition to remedying other Standard Model
deficiencies, provide a candidate dark matter particle. For example, dark matter candidates naturally arise in Supersymmetry, Kaluza-Klein models of extra dimensions and Little Higgs models (for a review see for example~\cite{Bertone04}). In this paper we consider dark matter candidates that arise in the Simplest Little Higgs with T-parity model (SLHT), as proposed by Martin~\cite{amartin}.

Various astronomical observations constrain properties of dark matter. The simplest constraint is on the amount of dark matter in the universe. Recent observations have improved measurements of the total mass density and baryonic mass density to the point that we now know the total amount of dark matter to a precision of 10$\%$~\cite{Spergel07}. The high precision of the current astrophysical searches allows to move beyond providing a general confirmation of the CDM paradigm to specifically constraining particle physics models based on the dark matter properties they predict. In the present study we focus on the relic density constraints, which provide the most significant reduction of the allowed parameter space for any extensions of the SM. 

Imposing astrophysical constraints leads to a set of specific predictions for the observation and analysis of dark matter at the Large Hadron Collider at CERN. We analyze collider signatures of the Simplest Little Higgs dark matter and discuss prospects and difficulties for dark matter detection at present day collider facilities. We see this analysis as a great example of synergy between particle physics and astronomy in solving one of the most outstanding problems in science. 

The rest of the paper is organized as follows. In Section~2 we review the foundations of the Simplest Little Higgs with T-parity (SLHT) model and discuss dark matter candidates that arise in it. In Section~3 we calculate the relic density constraints on the parameter space of the SLHT. Collider phenomenology of the SLHT dark matter is discussed in Section~4. A summary of our results follows in Section 5. Finally, in the Appendix, we provide Feynman rules for the Simplest Little Higgs with T-parity model.

\section{Simplest Little Higgs with T-parity}
Little Higgs models \cite{AH2001, AH2002, AH02, littlest, Low2002, Schmaltz2002} offer an interesting alternative to supersymmetry in solving various problems of the SM. For instance, the hierarchy problem is resolved by introducing new heavy particles at the TeV scale that cancel quadratic divergences of the Higgs boson mass arising in the SM from one-loop radiative corrections. This cancellation occurs between particles with the same statistics in Little Higgs models, as opposed to the case of supersymmetry where divergences are canceled by opposite statistics partners of the SM particles. The original little Higgs models~\cite{littlest,KS,Schmaltznote}, however, suffered from stringent constraints from electroweak precision tests (EWPT) which required the introduction of fine tuning. An elegant solution to this problem is to introduce a discrete symmetry (like the R-parity in supersymmetry) called T-parity~\cite{LHT2}. T-parity not only remedies the problem of precision electroweak constraints, but also leads to the appearance of promising dark matter candidates. We must note that, recently, it has been pointed out that T-parity in general may be violated by anomalies~\cite{Hill}, making the lightest T-odd particle unstable. However, such violation strongly depends on the UV completion of the Little Higgs model and may not be realized in nature~\cite{csaki}. Consequently, we consider models respecting T-parity here. 

The Simplest Little Higgs with T-parity model~\cite{amartin} is constructed by enlarging the SM $SU(2)_L \otimes U(1)_Y$ gauge group to $SU(3)_W\otimes U(1)_X$ in a minimal way.  This entails enlarging the SU(2) doublets of the SM to SU(3) triplets, including additional SU(3) gauge bosons, and writing SU(3) invariant interactions in a way that reproduces all the SM couplings when restricted to SM fields. The $SU(3)_W\otimes U(1)_X$ gauge symmetry is broken down to the SM electroweak gauge group by two complex scalar fields $\Phi_{1,2}$, which are triplets under the SU(3) with aligned vevs $f_{1,2}$. When the scalar fields acquire vevs, the initial global symmetry $[SU(3)_W\otimes U(1)_X]^2$ is spontaneously broken to $[SU(2)\otimes U(1)_X]^2$. At the same time, the global symmetry is explicitly broken to its diagonal subgroup $SU(3)\otimes U(1)$ by the gauge interactions. 

The scalar fields are conveniently parameterized by a nonlinear sigma model of the $[SU(3)\otimes U(1)_X]^2/[SU(2)\otimes U(1)_X]^2$ symmetry breaking as
\begin{eqnarray}
\Phi_1= e^{i\Theta  \frac{f_2}{f_1} }
\left( \begin{array}{l}
0  \\ 0 \\ f_1 \end{array} \right) , \quad
\Phi_2= e^{-i \Theta \frac{f_1}{f_2}}
\left( \begin{array}{l}
0  \\ 0 \\ f_2\end{array} \right) 
\end{eqnarray}
where
\begin{eqnarray}
\Theta = \frac{1}{f}\left[
\frac{\eta}{\sqrt{2}} +
\left( \begin{array}{cc} 
\!\!\begin{array}{ll} 0 & 0 \\ 0 & 0 \end{array} 
& \!\ \\ h^\dagger  & \!\!0 \end{array} \right)\right]
 \quad  {\rm and}\quad  f^2 = f_1^2+f_2^2 \ .
\label{eq:phiexpand}
\end{eqnarray}

Here $\eta$ is a real scalar field and $h = (h^0,h^-)^T$ is an $SU(2)$ doublet which can be identified as the SM Higgs. The remaining 5 degrees of freedom of the symmetry breaking are ``eaten'' by the additional heavy gauge bosons and therefore are omitted from Eq.(\ref{eq:phiexpand}). 

Scalar field interactions with the gauge boson sector and the masses of gauge bosons are determined by the standard kinetic term for the $\Phi_i$~\cite{Hanslht}:
\begin{equation}
	\mathcal{L}_{\Phi} = \left| \left( \partial_{\mu}
	+ i g A_{\mu}^a T^a - \frac{i g_x}{3} B_{\mu}^x \right) 
	\Phi_i \right|^2,
	\label{SU3LPhi}
\end{equation}
where the SU(3) gauge coupling $g$ is equal to the SM SU(2)$_L$ gauge coupling
and the U(1)$_X$ gauge coupling $g_x$ is fixed in terms of $g$ and 
the weak mixing angle $t_W \equiv \tan\theta_W$ by
\begin{equation}
	g_x = \frac{g t_W}{\sqrt{1 - t_W^2/3}}.
\end{equation}

The gauge bosons corresponding to the broken generators get masses of order $f \sim$ TeV and consist of a $Z_h$ boson, which is a linear combination of $A^8$ and $B^x$
\begin{equation}
Z_h = \frac{1}{\sqrt{3}}\left( \sqrt{3 - t_W^2} A^8 + t_W B^x \right),
\end{equation}
and a complex SU(2)$_L$ doublet $(Y^0,X^-)$
\begin{equation}
X^- = \frac{A^6 - iA^7}{\sqrt{2}},  \ \ \ \  Y^0 = \frac{A^4 - iA^5}{\sqrt{2}}.
\end{equation}
The SM mass eigenstates $W^{\pm}$, $Z^0$, and $A$ are also expressed in terms of the gauge fields:
\begin{equation}
W^{\pm} = \frac{A^1 \mp iA^2}{\sqrt{2}},
\end{equation}
\begin{equation}
Z^0 = A^3 c_W + \frac{t_W}{\sqrt{3}}(A^8 s_W - B^x c_W \sqrt{3 - t_W^2}),
\end{equation}
\begin{equation}
A = A^3 s_W - A^8\frac{s_W}{\sqrt{3}} + B^x\frac{c_W}{\sqrt{3}}\sqrt{3 - t_W^2}.
\end{equation}
As in the SM, masses of the heavy gauge bosons are determined by the symmetry breaking vev and can be written as:
\begin{equation}
M_{Z_h} = \frac{\sqrt{2} g f}{\sqrt{3 - t_W^2}},\ \ \ \ M_{X^{\pm}} = M_{Y^0} = \frac{g f}{\sqrt{2}}.
\label{eq:slhtmasses}
\end{equation}

The symmetry breaking $SU(2)\times U(1)/U(1)_{em}$ by the vev of the SM Higgs introduces corrections to the masses of the gauge bosons. For example, the charged heavy bosons $X^{\pm}$ become slightly lighter than the neutral gauge boson $Y^0$. However, for our studies of dark matter physics, we are interested only in the first order effects and therefore will disregard $(v/f)^2$ corrections to masses and to the couplings in the Feynman rules. 

There are two possible gauge charge assignments for fermions in the Simplest Little Higgs model. The first (universal embedding) yields heavy partners of the top, charm and up quarks, with all generations carrying identical quantum numbers. However, this embedding is known to be anomalous and also appears to be ruled out by precision electroweak constraints~\cite{Schmaltznote}. In the second scenario (anomaly-free embedding) the top, strange and down quarks have heavy partners. The first two generations of quarks transform under the antifundamental $\mathbf{\bar{3}}$ representation of the SU(3), while the third generation quarks and leptons transform as $\textbf{3}$'s~\cite{Hanslht}:
\begin{eqnarray}
  L^T_m = (\nu, e, iN)_m, \qquad \qquad   ie^c_m, iN^c_m,
  \nonumber \\
  Q^T_1 = (d, -u, iD),\quad \qquad  id^c, iu^c, iD^c,
	\nonumber \\
	Q^T_2 = (s, -c, iS),\;  \quad \qquad  is^c, ic^c, iS^c,
	\nonumber \\
	Q^T_3 = (t,b,iT), \; \qquad \qquad   it^c, ib^c, iT^c,	
\end{eqnarray} 
where $m$ is the generation index, and $u^c$, $d^c$, ... are the right-handed Weyl fermions invariant under $SU(3)_W$ that marry corresponding components of the $SU(3)_W$ triplets to produce fermion masses. Hypercharge assignments and representations of the fundamental fields in the anomaly-free embedding are summarized in Table~\ref{slhtrpn}. We consider only this case.   

\begin{table}
\centering
\begin{tabular}{|c||c|}
\hline
 & Anomaly-free embedding \\
\hline
fermion & $(SU(3)_c \times SU(3)_w)_{U(1)_X}$ \\
\hline
$Q_{1,2}$ & $(\mathbf{3},\ \mathbf{\bar{3}})_{0}$ \\
\hline
$Q_3$ & $(\mathbf{3},\ \mathbf{3})_{\frac{1}{3}}$ \\
\hline
$u^c_a, T^c$ & $(\mathbf{\bar{3}},\ \mathbf{1})_{\frac{-2}{3}}$ \\
\hline
$d^c_a, D^c, S^c$ & $(\mathbf{\bar{3}},\ \mathbf{1})_{\frac{1}{3}}$ \\
\hline
$L_m$ &  $(\mathbf{1},\ \mathbf{3})_{\frac{-1}{3}}$ \\
\hline
$N^c_m$ &  $(\mathbf{1},\ \mathbf{1})_{0}$ \\
\hline
$e^c_m$ & $(\mathbf{1},\ \mathbf{1})_{1}$ \\
\hline
\end{tabular}
\caption{\label{slhtrpn} The $SU(3)_c\otimes SU(3)_W \otimes U(1)_X$ representations of the fermions in the anomaly-free embedding of the Simplest Little Higgs with T-parity model. Index $a$ runs over the first two generations of quarks, while $m$ runs over all three generations of leptons.}
\end{table}

Following the initial investigation~\cite{amartin}, we further introduce a discrete symmetry (T-parity) into the model to avoid precision electroweak constraints and to produce a heavy stable particle as a dark matter candidate. A straightforward way to implement T-parity into the Simplest Little Higgs model is to start with the fermionic sector. The action of the T-parity operator on the fermion triplets can be defined as~\cite{amartin}
\begin{equation}
\Psi_{Q,L} \rightarrow -\hat{\Omega} \Psi_{Q,L},
\end{equation}
where $\hat{\Omega} = diag(-1,-1,1)$. For the scalar fields the transformation is
\begin{equation}
\Phi_1 \rightarrow \hat{\Omega}\Phi_2 
\end{equation}
which immediately requires $f_1 = f_2 = f/\sqrt{2}$. This definition of T-parity also determines the following transformation rules for the gauge boson sector, which follow from the requirement of the T-parity invariance for the kinetic term~(\ref{SU3LPhi}),
\begin{equation}
A_{\mu} \rightarrow \hat{\Omega}A_{\mu}\hat{\Omega}.
\end{equation}

It is important to note that, under this transformation, the new neutral gauge boson $Z_h$ remains even, and hence its mass is required to be relatively heavy in order to avoid the precision electroweak constraints.

T-parity invariance requires that both scalar triplets $\Phi_1$ and $\Phi_2$ couple to the fermion fields with equal strength. This introduces major modifications to the standard Simplest Little Higgs model. Yukawa interactions in the leptonic sector can be written as
\begin{eqnarray}
	\mathcal{L}_l = i \lambda_{N_m} N_m^c(\Phi^{\dag}_1 + \Phi^{\dag}_2)L_m + \\ \nonumber
	+ \frac{i \lambda_e^m}{\Lambda} e_m^c \epsilon_{ijk} 
	\Phi_1^i \Phi_2^j L_m^k + {\rm h.c.},
\end{eqnarray}
where $m = 1,2,3$ is a generation index and $i,j,k = 1,2,3$ are SU(3) indices. Yukawa Lagrangians for the third generation of quarks and for the first two generations are, respectively,
\begin{eqnarray}
\mathcal{L}_{3} = i\lambda_t Q_3[t_L^c\left( \Phi_2^+ - \Phi_1^+ \right) + \qquad \\ \nonumber
+ t_H^c \left( \Phi_2^+ + \Phi_1^+ \right)] + \frac{\lambda_b}{\Lambda} ib^c \epsilon_{ijk} \Phi_1^i \Phi_2^j Q_3^k + {\rm h.c.},
\end{eqnarray}
\begin{eqnarray}
\mathcal{L}_{n} = i \lambda_D^n Q_n^T (\Phi_1 + \Phi_2)d_H^{nc} + \qquad \\ \nonumber
		+ i \lambda_d^n Q_n^T (\Phi_1 - \Phi_2)d_L^{nc} + i\frac{\lambda_u^{n}}{\Lambda}
		u^c_n \epsilon_{ijk} \Phi_1^{*i} \Phi_2^{*j} Q_n^k + {\rm h.c.},
\end{eqnarray}
where $n = 1,2$; $i,j,k = 1,2,3$ are SU(3) indexes; $t_L^c$ and $t_H^c$ are T-even and T-odd linear combinations of $t^c$ and $T^c$, while $d^{nc}_H$ and $d^{nc}_L$ are linear combinations of $d^c$ and $D^c$ for $n=1$ and of $s^c$ and $S^c$ for $n=2$; $u^c_m$ runs over all the up-type conjugate quarks ($u^c, c^c, t^c$). Also, we are neglecting flavor mixing here.

The requirement of T-invariance leads to a simple relation for the heavy top mass: 
\begin{equation}
M_T = M_t\frac{\sqrt{2}f}{v}.
\end{equation}
The masses of the remaining new fermions come from the Yukawa interaction in the form $M_f  = \sqrt{2}\lambda_f f$. We choose a universal value of the Yukawa coupling $\lambda$ for all new fermions, except for the first generation heavy neutrino. In this work we choose $\lambda$ to be large and hence additional fermions decouple and have negligible influence on the results considered. We also note that a more general study without decoupling of the fermions should be undertaken. 

Introduction of T-parity makes the lightest T-odd particle (LTP) stable. There are two neutral heavy particles that should be considered as the potential LPT: a heavy neutrino $N$ and a new scalar $\eta$. The mass of $\eta$ depends on the mass of the SM Higgs boson as well as on the masses of other heavy particles through the Coleman-Weinberg potential and is generally of order $M_{\eta} \gtrsim 700$ GeV~\cite{Hanslht}. Although it is possible to make $\eta$ a LTP it requires significant fine tuning in the parameter space. Moreover, the coupling of $\eta$ to other particles in the primordial plasma is usually $f^2$ suppressed which makes it practically impossible to satisfy relic density constraints~\cite{amartin}. On the other hand, the mass of a heavy neutrino remains a free parameter in the model and can be chosen to be significantly lighter than the masses of other particles in the model. A heavy neutrino is weakly coupled to the rest of the particles through $Z_h$ and $X^{\pm}$, $Y$ exchanges and it decouples at the right time to produce the dark matter density observed today. In  light of this, we choose to consider a heavy neutrino as a dark matter candidate of the Simplest Little Higgs with T-parity Model.

\section{Relic abundance constraints}
The amount of dark matter today, the relic density, has been measured with good precision by WMAP as $0.094 < \Omega_{DM} h^2 < 0.126$~\cite{Spergel07}. Here $\Omega_{DM}$ is the ratio of dark matter density to the critical density of the universe and $h$ parametrizes the Hubble constant today. From a theoretical side we can calculate the relic density of dark matter in terms of the major parameters of any models that produce a stable WIMP and hence impose stringent constraints on the allowed parameter space of the models. 

In the early universe, dark matter particles were present in large numbers in thermal equilibrium with the rest of the cosmic plasma; as the universe cooled, they could reduce their density through pair annihilation. As their density decreases, however, it becomes more and more difficult for particles to find partners to annihilate with, and the comoving density becomes constant, known as ``freeze-out''. Therefore, the present day density of dark matter particles is completely determined at the time of freeze-out. It can be conveniently expressed in terms of the thermally averaged product of the annihilation cross velocity $\langle\sigma v \rangle$ at the time of freeze-out~\cite{Kolb}: 
\begin{equation}
\label{eq:relic}
\Omega_{\chi} h^2 \approx \frac{1.07 \times 10^9 \, \rm{GeV}^{-1}}
{M_{Pl}}\frac{x_F}{\sqrt{g_*}}\frac{1}{\langle\sigma v\rangle}.
\end{equation}
Here $g_*$ counts the number of relativistic degrees of freedom and is approximately equal to the number of bosonic relativistic degrees of freedom plus $7/8$ of the fermionic relativistic degrees of freedom  evaluated at the freeze-out temperature (a plot of $g_*$ as a function of time can be found for instance in \cite{Kamionkowski}). Also $x_F \equiv m/T_F$ parametrizes freeze-out temperature $T_F$ and can be estimated through iterative solution of the equation 
\begin{equation}
\label{XF}
x_F=\ln \left[c (c+2) \sqrt{\frac{45 }{8}}
\frac{g}{2\pi^3} \frac{m \ M_{Pl} \langle\sigma v\rangle}{g^{1/2}_* x^{1/2}_F}\right],
\end{equation}
where $c$ is a constant of order one determined by matching the late-time and early-time solutions.

\begin{figure*}
\centering
\includegraphics{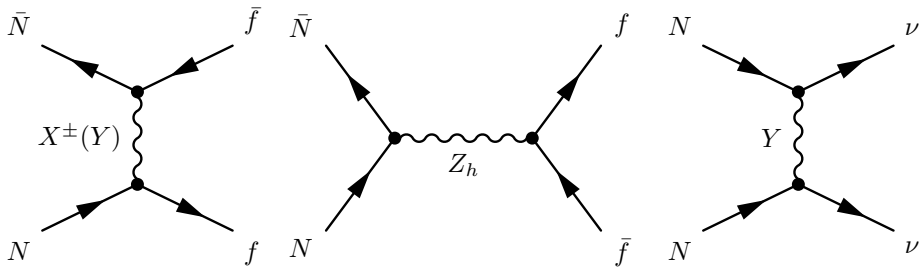}
\caption{\label{slhtannih}Dominant channels for dark matter annihilation in the Simplest Little Higgs with T-parity model. These diagrams provide the largest contributions to the heavy neutrino annihilation cross section for the range of SLHT parameters examined in this paper.}
\end{figure*}

\begin{figure}
\includegraphics[width=3.4in]{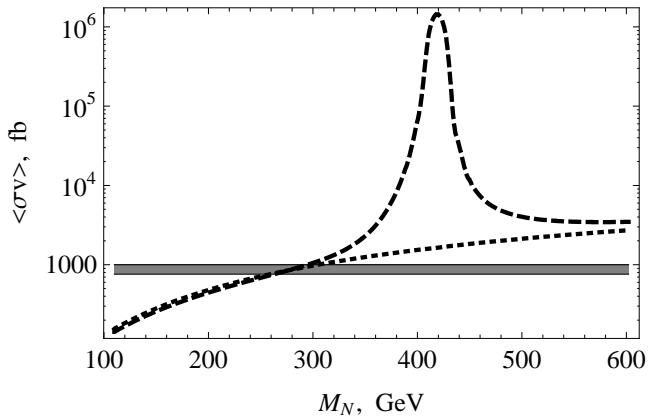}
\caption{\label{slhtcrossx}t-channel contribution (short-dash line) to the thermally averaged annihilation cross section $\left\langle \sigma v\right\rangle$ (dashed line) in the Simplest Little Higgs with T-parity model as a function of dark matter mass. The WMAP allowed region (gray) is also shown for reference. This plot emphasizes the contribution of the t-channel diagrams to the annihilation cross section.}
\end{figure}

Major annihilation reactions that kept the heavy neutrino coupled to the primordial plasma before decoupling are shown in Fig.~\ref{slhtannih}. It is important to emphasize that in this paper we consider additional quarks as well as the second and third generation heavy neutrinos to be much heavier than the lightest T-odd particle and, hence, the effects of coannihilations are ignored. Without coannihilations the dark matter relic density is determined by the $Z_h$ s-channel exchange as well as by the t-channel $X^{\pm}$ and $Y$ processes. The t-channel processes are especially important for the heavy LTP. The increase in the cross section caused by the t-channel processes and its impact on the dark matter freeze-out is depicted in Fig.~\ref{slhtcrossx}.

\begin{figure}
\centering
\includegraphics[width = 3.4in]{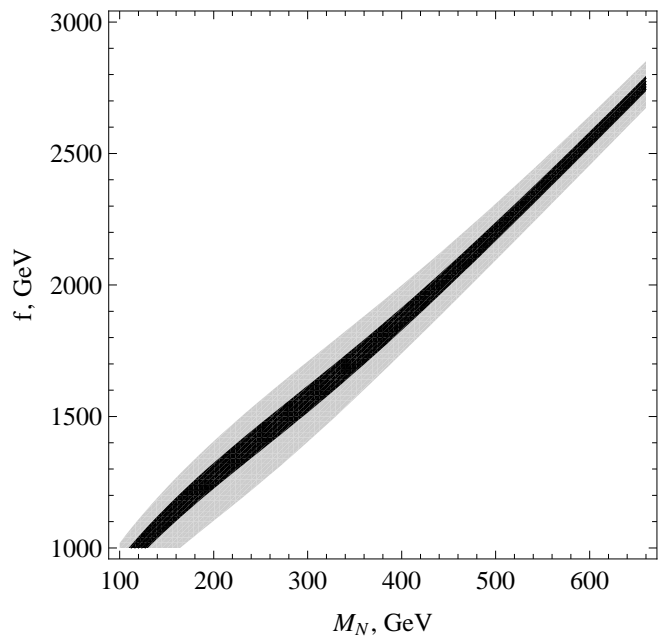}
\caption{\label{slhtconstraints}Relic abundance constraints on the parameter space of the Simplest Little Higgs with T-parity model. On this plot the dark region corresponds to 2$\sigma$ variation about the WMAP observed relic density, while the gray region covers 5$\sigma$ confidence level.}
\end{figure}

For the calculation of the relic density constraints we wrote a Mathematica code using the Feyncalc package to calculate invariant amplitudes of the annihilation processes shown in Fig.~\ref{slhtannih}. The constraints that arise are provided in Fig.~\ref{slhtconstraints}. We plot allowed regions that fall within 2$\sigma$ and 5$\sigma$ around the result obtained by the WMAP team $0.094 < \Omega_{DM} h^2 < 0.126$. The allowed region can be conveniently parametrized by a linear relation that links the dark matter mass, $M_N$, and the symmetry breaking scale, $f$, of the initial global symmetry in the SLHT:
\begin{equation}
\label{eqconstr}
f = 3.2M_N + 600,
\end{equation}
where $f$ and $M_N$ are in GeV. 

\section{Collider signatures}
As we have shown in the previous sections, astrophysical observations provide stringent constraints on the properties of dark matter particles and, hence, can direct the search for dark matter at existing and future colliders. Astrophysical probes, however, are unable to determine and study the exact properties of a dark matter particle and it largely remains for collider physics to unravel the mystery surrounding dark matter. 

Regardless of its origin, if cold dark matter is composed of WIMPs, then it may be possible to produce and study the dark matter particle(s) directly at the LHC~\cite{Baer} . In any collider experiment, WIMPs would be like neutrinos in that they would escape the detector without depositing any energy in the experimental apparatus, resulting in an apparent imbalance of energy and momentum in collider events. While WIMPs would manifest themselves only as missing (transverse) energy  at collider experiments, it should nevertheless be possible to analyze the visible particles produced in association to study the new physics associated with the WIMP sector. 

\begin{figure*}
\centering
\includegraphics{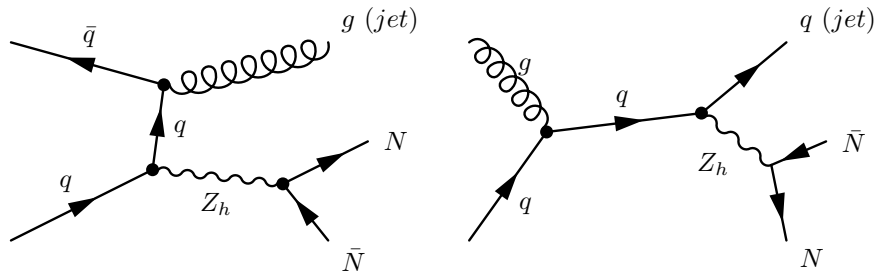}
\caption{\label{monojet}Sample Feynman diagrams for the monojet associated dark matter production at the LHC in the Simplest Little Higgs with T-parity model.}
\end{figure*}

\begin{figure*}
\centering
\includegraphics{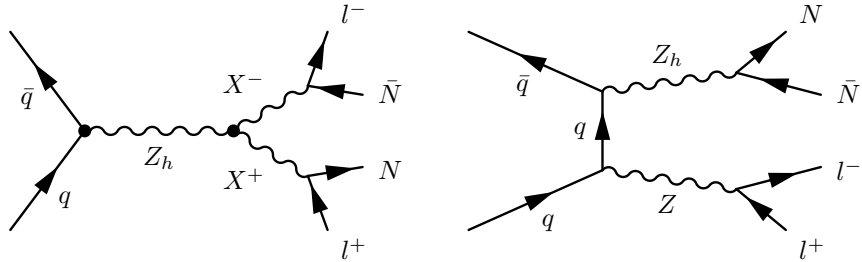}
\caption{\label{dilepton}Sample Feynman diagrams for the dileptonic dark matter production at the LHC in the Simplest Little Higgs with T-parity model.}
\end{figure*}

\begin{figure}
\centering
\includegraphics[width = 3.4in]{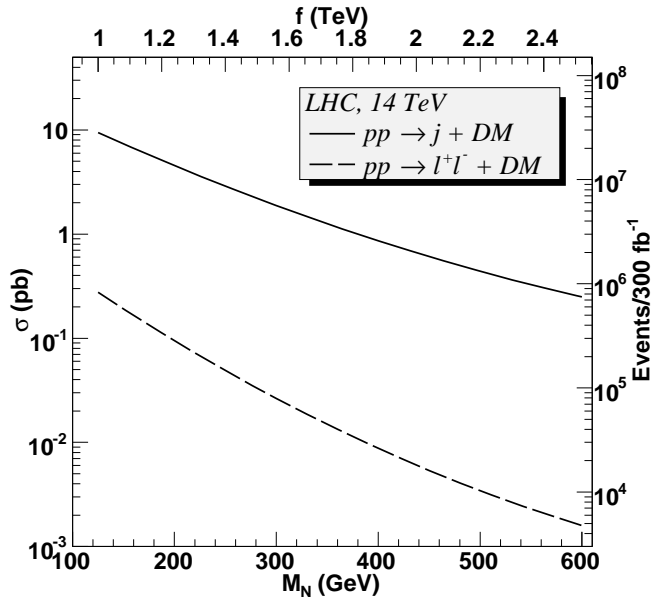}
\caption{\label{slhtCrX}Cross section of dark matter production at the LHC is plotted as a function of dark matter mass. On the top scale we identify the SLHT symmetry breaking scale $f$ that corresponds to the dark matter mass via the relic density constraints of Eq.~\ref{eqconstr}. The number of events for an integrated luminosity of 300 $fb^{-1}$ is plotted on the second y-axis.}
\end{figure}

We identify two promising channels of dark matter production for the model under consideration: 
\begin{center}
$pp \rightarrow jet + N \bar{N}$,
\end{center}
\begin{center}
$pp \rightarrow l^+ l^- + N \bar{N}$,
\end{center}
where $l^+ l^-$ represents an $e^+ e^-$ or $\mu^+ \mu^-$ pair. Feynman diagrams contributing to these channels are presented in Figs.~\ref{monojet} and \ref{dilepton}, respectively. Cross sections at LHC energy as a function of dark matter mass for these two channels are shown in Fig.~\ref{slhtCrX}. Monojet associated dark matter production is enhanced by the QCD coupling and we can expect a large number of events for this channel. A large cross section of the monojet plus missing energy channel makes it an extremely important signature of the Simplest Little Higgs with T-parity model. This signature may help to distinguish this model from both supersymmetry, in which dark matter is predominantly produced in multijet events, and the Littlest Higgs with T-parity model, in which monojet dark matter production is highly suppressed. In fact we identify monojet associated dark matter production as a smoking-gun signature of the Simplest Little Higgs with T-parity. If the SLHT is realized in nature then this process should be observed at a very early stage of the LHC operation and the model can be confirmed with the dilepton channel which has a distinct signature, but a lower cross section.    

\begin{figure*}
\centering
\includegraphics{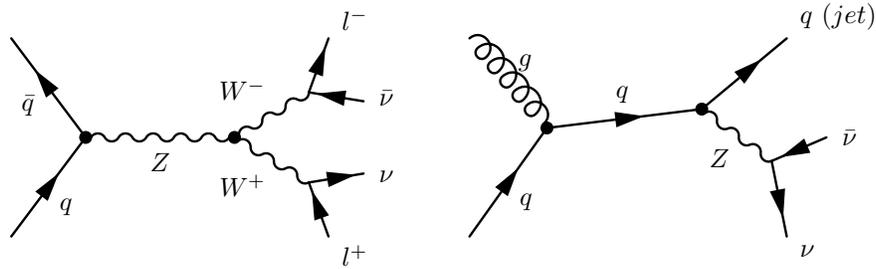}
\caption{\label{smbg}Sample Feynman diagrams for the Standard Model background to the dileptonic (left) and monojet (right) associated dark matter production at the LHC.}
\end{figure*}

The integrated luminosity of the LHC is expected to be around $10\ fb^{-1}$ after the first year of operation and more than $300\ fb^{-1}$ over the full period of operation. From Fig.~\ref{slhtCrX} we can see that, if the Simplest Little Higgs mechanism is realized in nature, and the heavy neutrino is light enough, then dark matter indeed will be copiously produced at the LHC. However, the SM model background is also significant. Some Feynman diagrams contributing to the SM background of the monojet and dilepton channels are presented in Fig.~\ref{smbg}. The total cross section at LHC for the monojet plus missing energy production in the SM is $\sigma \approx 2\cdot10^3\ pb$, while for the dilepton plus missing energy channel $\sigma \approx 2\ pb$. In both cases we expect many more background events than DM events. However, because of the large mass of the dark matter particle, signal events appear in the kinematic regions where the SM background is relatively small. 

To study the LHC signal of dark matter production in the SLHT we implemented the model into the MadGraph package~\cite{madgraph}. Based on the relic density constraint we have chosen three sets of parameters that cover most of the allowed parameter space and hence are representative of the model in general.Our parameters are given in Table~\ref{slhtparams}. Also, for all collider calculations we use the standard set of major SM parameters \cite{pdg}: $G_F = 1.166 37(1)\times10^{-5}$ GeV$^{-2}$ -- Fermi constant, $\alpha~=~1/137.036(1)$ -- fine-structure constant, $sin\theta_W^2(M_Z) = 0.23152(14) $ -- square of the electroweak mixing angle, and $\alpha_s(M_Z) = 0.1176(20)$ -- strong coupling constant. 

\begin{figure}
\centering
\includegraphics[width = 3.4in]{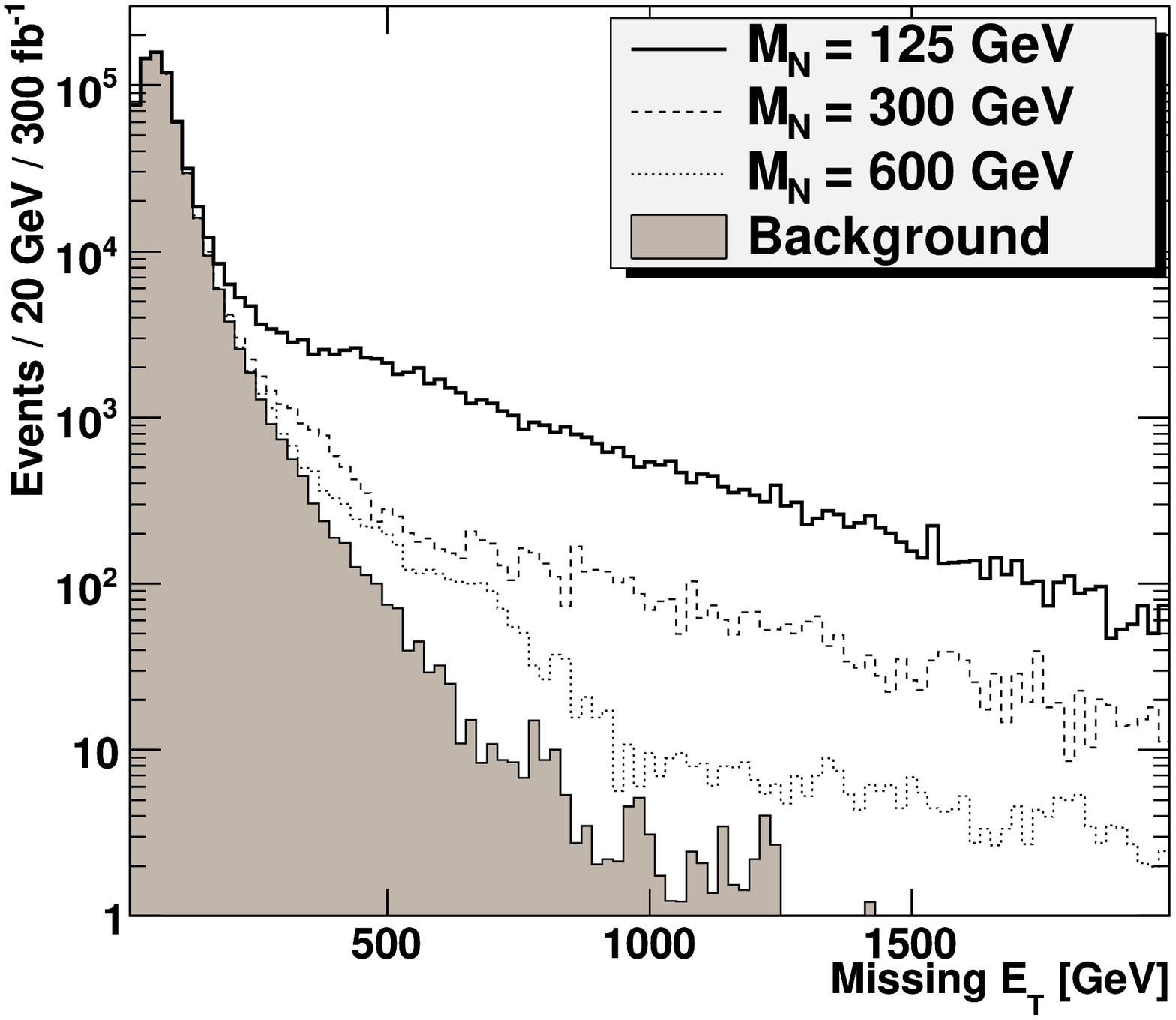}
\caption{\label{ll1}Missing energy distribution expected in the SLHT in the dilepton plus missing energy channel. The SM background is also plotted for comparison.}
\end{figure}

\begin{figure}
\centering
\includegraphics[width = 3.4in]{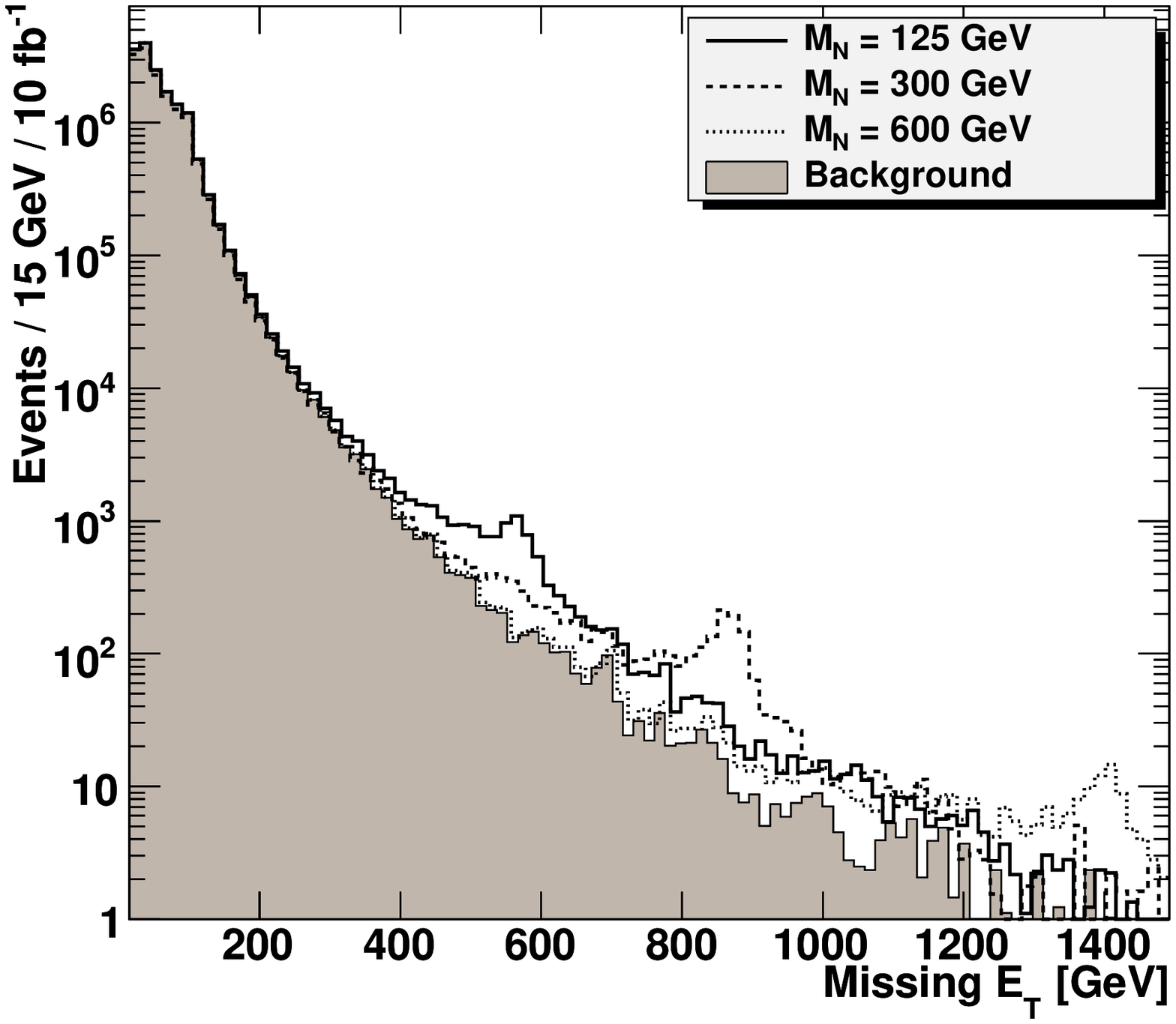}
\caption{\label{j1}Missing energy distribution expected in the SLHT in the one jet plus missing energy channel. The SM background is also plotted for comparison.}
\end{figure}

In Fig.~\ref{ll1} and~\ref{j1} we show the predicted signal of the two missing energy channels compared to the SM background for various values of the dark matter mass. A bin-by-bin $\chi^2$ analysis of the data clearly shows the possibility of 5$\sigma$ detection of dark for the considered region of SLHT parameter space.  

\begin{table}
\centering
\vspace{20pt}
\begin{tabular}{|c||c|c|c|}
\hline
Parameter in GeV & Set 1 & Set 2 & Set 3 \\
\hline
f & 1000  & 1560 &  2520 \\
\hline
$M_{N}$ & 125 & 300  & 600 \\
\hline
$M_{Z_h}$ & 560 & 872 & 1409 \\
\hline
$M_h$ & 120 & 120 & 120 \\
\hline
$\Gamma_{Z_h}$ & 4.0 &  5.8 & 8.6 \\
\hline
$M_X$ & 460  & 717 & 1158 \\
\hline
$\Gamma_X$ & 1.1 & 1.5 & 2.0 \\
\hline
$M_T$ & 983 & 1533 & 2477 \\
\hline
$M_D,\ M_S,\ M_{N_2},\ M_{N_3}$ & 707 & 1103 & 1782 \\
\hline
\end{tabular}
\caption{\label{slhtparams}Parameters used for the analysis of dark matter production at the LHC in the Simplest Little Higgs with T-parity model. Values of the tested parameters are chosen to be consistent with the relic density constraints from Fig.~\ref{slhtconstraints} and to cover most of the allowed parameter space.}
\vspace{20pt}
\end{table}

\begin{figure}
\centering
\includegraphics[width = 3.4in]{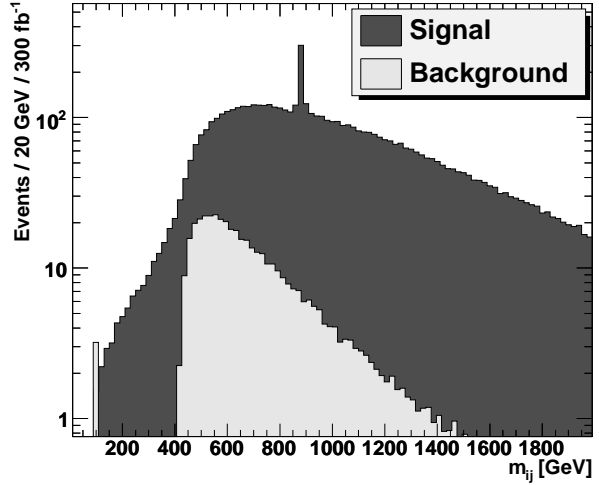}
\includegraphics[width = 3.4in]{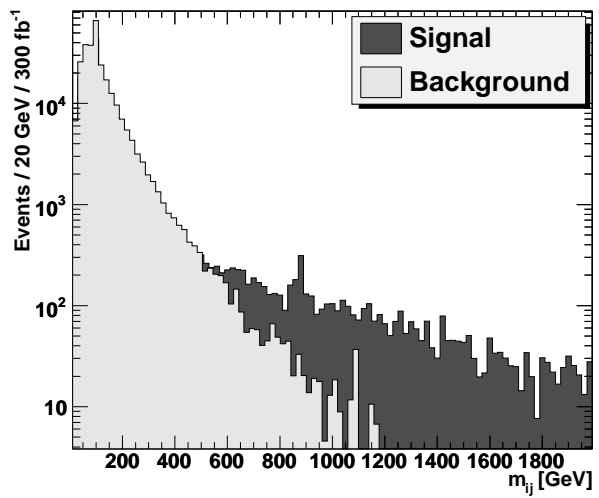}
\caption{\label{mijcut}Invariant mass distribution of electrons in dilepton plus missing energy production for $M_N = 300$ GeV with $p_T > 200$ GeV cut imposed(top) and without cuts (bottom). SM expectation for this channel is also presented.}
\end{figure}

In order to increase the signal to background ratio we can impose kinematic cuts on the observed particles as well as on the missing energy. To determine optimal cuts we considered predicted distributions for the observed particles as well as for missing energy in both the SLHT and the SM. In this study we analyze cuts only for the dilepton channel, since the monojet signal is very well distinguished over the SM background without introducing any cuts over all values of the considered parameters. Leptons produced in the dark matter related processes are more energetic and mainly appear in the central region of the detector. This suggests the introduction of a variety of kinematic cuts depending on the specific objectives. In this paper, we considered a cut on the charged lepton momentum $p_T$ which allows us to significantly increase the threshold on the dark matter mass that can be observed in the dilepton channel. In Fig.~\ref{mijcut} we show the dilepton plus missing energy signal for the heavy neutrino mass of $M_N = 300$ GeV both before and after kinematic cut $p_T > 200$ GeV. This cut reduces the cross section of the SM background by more than 3 orders in magnitude, while the signal is decreased only 2 orders. Remarkably, this cut also allows us to study a different aspect of the SLHT model, namely the mass of the new heavy neutral gauge boson $Z_h$, with the observation of the invariant mass distribution for the charged leptons. In Fig.~\ref{mijcut}, we clearly see that the distribution in the SLHT model has a distinctive peak around the mass of $Z_h$ that is strongly emphasized by the introduction of the cut.

\section{Summary}
In this paper, we have investigated dark matter candidates that arise in the Simplest Little Higgs model after the introduction of a discrete $Z_2$ symmetry called T-parity. A heavy neutrino is identified as the most promising dark matter candidate of this model and relic density constraints on the model's parameter space are imposed. It has been shown that, if the heavy neutrino is responsible for the observed dark matter in the universe, then its mass is approximately linearly related to the value of the SLH symmetry breaking scale $f$ as $f \simeq 3.2M_N + 600$, where $f$ and $M_N$ are in GeV.

We have analyzed couplings specific to the Simplest Little Higgs with T-parity (SLHT) model and introduced the model into the MadGraph package.  
Collider phenomenology of SLHT dark matter was discussed and we have shown that, for a large region of the allowed parameter space, dark matter particles could be detected at the LHC. Problems of the dark matter analysis at the LHC were outlined. We showed that the SLHT has an important monojet plus missing energy signature that clearly distinguishes it from both the Littlest Higgs with T-parity model and from supersymmetry. A dileptonic channel of dark matter production was shown to be a promising independent channel of heavy neutrino production that can be used to confirm the SLHT mechanism if the larger monojet dark matter signal is detected. We also presented various kinematic distributions for the leptons of the dilepton associated dark matter production and discussed kinematic cuts that provide significant enhancement of the signal to background ratio and thus permit observation of dark matter particles with higher mass as well as allow to study various properties of the gauge sector of the SLHT. 
\begin{acknowledgments}
D.T. would like to thank Aliaksei Charnukha for valuable discussions and technical support throughout this research. D.T. is grateful to Dr. Michael Santos for his support and advice during this work. The authors are also grateful to Ken Moats for insightful discussions during the implementation of the SLHT into MadGraph and subsequent analysis of the data.  This research was supported in part by the Natural Sciences and Engineering Research Council of Canada. 
\end{acknowledgments}

\appendix*
\section{Feynman Rules for the Simplest Little Higgs with T-parity model}
In this appendix we provide a list of Feynman rules specific to the SLHT model. As discussed in Section 2, implementing T-parity in the gauge and scalar sectors of the Simplest Little Higgs model does not require significant change to the structure of the original Simplest Little Higgs model. The major effect in the gauge sector is to set the aligned vevs of the two global $[SU(3)_W \otimes U(1)_X]_i$ symmetry factors equal, $f_1 = f_2$. In the notation of~\cite{Hanslht}  this requirement leads to $tan\beta=f_2/f_1=1$. We summarize the triple and quartic gauge boson vertex factors in Tables~\ref{triplegb} and~\ref{quarticgb}, respectively. The triple and quartic couplings are given in the form
\begin{eqnarray}
\nonumber
V_1^{\mu}(k_1)V_2^{\nu}(k_2)V_3^{\rho}(k_3) : -i [ g^{\mu\nu} (k_1 - k_2)^{\rho} +   \\ 
\nonumber
+ g^{\nu\rho} (k_2 - k_3)^{\mu} + g^{\rho\mu} (k_3 - k_1)^{\nu} ] g_{V_1V_2V_3},
\end{eqnarray}
\begin{eqnarray}
\nonumber
V_1^{\mu}V_2^{\nu}V_3^{\rho}V_4^{\sigma} : \qquad \qquad \qquad \\
\nonumber
- i (2g^{\mu\nu}g^{\rho\sigma} - g^{\mu\rho}g^{\nu\sigma} - g^{\nu\rho}g^{\mu\sigma}) g_{V_1V_2V_3V_4},
\end{eqnarray}
where $g_{V_1V_2V_3}$ and $g_{V_1V_2V_3V_4}$ are vertex factors provided in the tables. We also must note that the $W^-Z^0X^+Y^0$ and $W^+Z^0X^-\bar{Y}^0$ couplings can not be written in the standard form due to the approximate nature of the gauge eigenstates. The appropriate couplings have the form: 

\begin{eqnarray}
\nonumber
W_{\mu}^-Z_{\nu}^0X_{\rho}^+Y_{\sigma}^0 : \quad \frac{-is_W^2}{2\sqrt{2}c_W}[2g_{\mu\nu}g_{\rho\sigma}- \\
\nonumber
-g_{\mu\sigma}g_{\nu\rho} - g_{\mu\rho}g_{\nu\sigma}] - 3c_W^2(g_{\mu\sigma}g_{\nu\rho} - g_{\mu\rho}g_{\nu\sigma}),
\end{eqnarray} 

\begin{eqnarray}
\nonumber
W_{\mu}^+Z_{\nu}^0X_{\rho}^-\bar{Y}_{\sigma}^0 : \quad \frac{-is_W^2}{2\sqrt{2}c_W}[2g_{\mu\nu}g_{\rho\sigma}- \\
\nonumber
- g_{\mu\sigma}g_{\nu\rho} - g_{\mu\rho}g_{\nu\sigma}] - 3c_W^2(g_{\mu\sigma}g_{\nu\rho} - g_{\mu\rho}g_{\nu\sigma}).
\end{eqnarray} 

\begin{table}[h!]
\center{\begin{tabular}{|c|c||c|c|} \hline Particles & $g_{VVV}$ &
Particles & $g_{VVV}$ \\

\feyn {W^+ W^- A} {-g s_W} {X^+ X^- A} {-g s_W}
\feyn {W^+ W^- Z^0} {-g c_W} {X^+ X^- Z^0} {-\frac{g}{2}\frac{1-2s_W^2}{c_W}}
\feyn {W^+ W^- Z_h} {0} {X^+ X^- Z_h} {\frac{g}{2}\sqrt{3-t_W^2}}
\feyn {Y^0 \bar{Y}^0 Z^0} {-\frac{1}{2c_W}g} {Y^0 \bar{Y}^0 A} {0}
\feyn {Y^0 \bar{Y}^0 Z_h} {-\frac{g}{2}\sqrt{3-t_W^2}} {X^+ W^- Y^0} {-\frac{g}{\sqrt{2}}}

\hline
\end{tabular}}
\caption{\label{triplegb}Feynman rules for the triple gauge boson interactions in the Simplest Little Higgs with T-parity model. All momenta are defined as outgoing. The SM couplings of $W^{\pm}$ to $Z^0$ and photon are also presented for reference. In the table $s_W$, $c_W$, and $t_W$ stand for the sine, cosine and tangent of the electroweak mixing angle respectively.}
\end{table}

\begin{table*}
\center{\begin{tabular}{|c|c||c|c|} \hline Particles & $g_{VVVV}$ &
Particles & $g_{VVVV}$ \\
\feyn {W^+W^-A A} {g^2s_W^2} {W^+ W^+ W^- W^-} {-g^2}
\hline
\feyn {X^+ X^- A A} {g^2 s_W^2} {X^+ X^- Z^0 Z^0} {\frac{g^2}{4 c_W^2}(1-2 s_W^2)^2}
\hline
\feyn {X^+ X^- Z_h Z_h} {\frac{g^2}{4}(3-t_W^2)} {X^+ X^- A Z^0} {\frac{g^2 s_W}{2 c_W}(1-2 s_W^2)}
\hline
\feyn {X^+ X^- A Z_h} {-\frac{g^2 s_W}{2}\sqrt{3 - t_W^2}} {X^+ X^- Z^0 Z_h} {-\frac{g^2}{4 c_W} (1-2 s_W^2)\sqrt{3 - t_W^2}}
\hline
\feyn {Y^0 \bar{Y}^0 Z_h Z_h} {\frac{g^2}{4}(3-t_W^2)} {Y^0 \bar{Y}^0 Z^0 Z^0} {\frac{1}{4 c_W^2}g^2}
\hline
\feyn {Y^0 \bar{Y}^0 Z_h Z^0} {\frac{g^2}{4c_W}\sqrt{3-t_W^2}} {Y^0 \bar{Y}^0 Y^0 \bar{Y}^0} {g^2}
\hline
\feyn {W^-X^+ A Y^0} {\frac{s_W}{\sqrt{2}} g^2} {W^- Z_h X^+ Y^0} {\frac{g^2}{2\sqrt{2}}\sqrt{3 - t_W^2}}
\hline
\feyn {W^+X^- A \bar{Y}^0} {\frac{s_W}{\sqrt{2}} g^2} {W^+ Z_h X^- \bar{Y}^0} {\frac{g^2}{2\sqrt{2}}\sqrt{3 - t_W^2}}
\hline
\feyn {X^+\bar{Y}^0X^-Y^0} {-\frac{g^2}{2}} {X^+X^-X^+X^-} {g^2}
\hline
\feyn {W^+Y^0W^-\bar{Y}^0} {-\frac{g^2}{2}} {W^+X^+W^-X^-} {-\frac{g^2}{2}}
\hline \hline
$W_{\mu}^-Z_{\nu}^0X_{\rho}^+Y_{\sigma}^0$ & \multicolumn{3}{|c|}{$\frac{-is_W^2}{2\sqrt{2}c_W}{(2g_{\mu\nu}g_{\rho\sigma} - g_{\mu\sigma}g_{\nu\rho} - g_{\mu\rho}g_{\nu\sigma}) - 
3c_W^2(g_{\mu\sigma}g_{\nu\rho} - g_{\mu\rho}g_{\nu\sigma})}$} \\
\hline
$W_{\mu}^+Z_{\nu}^0X_{\rho}^-\bar{Y}_{\sigma}^0$ & \multicolumn{3}{|c|}{$\frac{-is_W^2}{2\sqrt{2}c_W}{(2g_{\mu\nu}g_{\rho\sigma} - g_{\mu\sigma}g_{\nu\rho} - g_{\mu\rho}g_{\nu\sigma}) - 
3c_W^2(g_{\mu\sigma}g_{\nu\rho} - g_{\mu\rho}g_{\nu\sigma})}$} \\
\hline
\end{tabular}}
\caption{\label{quarticgb}Feynman rules for the quartic gauge boson interactions in the Simplest Little Higgs with T-parity model. Sample SM couplings are also provided for reference.}
\end{table*}

\begin{table*}
\centering
\begin{tabular}{|c|c||c|c||c|c|}
\hline
Particles & Vertices & Particles & Vertices & Particles & Vertices \\
\hline
$\eta \bar{N}_m \nu_m$ & $-i\frac{vM_{N_m}}{2f^2}P_L$ & $\eta \bar{T} t$ & $-i \frac{M_T}{\sqrt{2}f}P_L$& $\eta \bar{D} d$ & $-i\frac{vM_{D}}{2f^2}P_L$ \\
\hline
$\eta \eta \bar{N}_m N_m$ & $\frac{M_{N_m}}{2f^2}$ & $\eta \eta \bar{T} T$ & $\frac{M_{T}}{2f^2}$ & $\eta \eta \bar{D} D$ & $\frac{M_{D}}{2f^2}$ \\  
\hline
$h \bar{N}_m N_m$ & $\frac{vM_{N_m}}{2f^2}$ & $h \bar{T} T$ & $\frac{vM_{T}}{2f^2}$ & $h \bar{D} D$ & $\frac{vM_{D}}{2f^2}$  \\
\hline
$h h \bar{N}_m N_m$ & $\frac{M_{N_m}}{2f^2}$ & $h h \bar{T} T$ & $\frac{M_{T}}{2f^2}$ &  $h h \bar{D} D$ & $\frac{M_{D}}{2f^2}$\\
\hline 
\end{tabular}
\caption{\label{Hll}Feynman rules for scalar-fermion couplings in the Simplest Little Higgs with T-parity model. $P_L = (1 - \gamma^5)/2$ is a projection operator.  Only the leading terms are given.}
\end{table*}

\begin{table*}
\centering
\begin{tabular}{|c|c||c|c|}
\hline
Particles & Vertices & Particles & Vertices \\
\hline
$Z^{0\mu} \overline e e: $ & $\frac{g}{2c_W}\gamma^{\mu}[(c_W^2 - s_W^2)P_L - 2 s_W^2P_R]$ & $Z_h^\mu \overline \nu \nu:$ & $-\frac{g}{c_W \sqrt{3-4s_W^2}}\gamma^{\mu}(\frac{1}{2} - s_W^2)P_L$ \\
\hline
$Z_h^\mu \overline t t:$ & $-\frac{g}{c_W \sqrt{3-4s_W^2}}\gamma^{\mu}[(\frac{1}{2} - \frac{1}{3}s_W^2)P_L + \frac{2}{3}s_W^2P_R]$ & $Z_h^\mu \overline{N} N:$ & $-\frac{g}{c_W \sqrt{3-4s_W^2}}\gamma^{\mu}(-1 + s_W^2)P_L$  \\
\hline		
$Z_h^\mu \overline b b:$ & $-\frac{g}{c_W \sqrt{3-4s_W^2}}\gamma^{\mu}[(\frac{1}{2} - \frac{1}{3}s_W^2)P_L - \frac{1}{3}s_W^2P_R]$ & $X^-_{\mu} \overline{D} u:$  & $\frac{ig}{\sqrt{2}} \gamma_{\mu} P_L $ \\
\hline
$Z_h^\mu \overline u u:$ & $-\frac{g}{c_W \sqrt{3-4s_W^2}}\gamma^{\mu}[(-\frac{1}{2} + \frac{2}{3}s_W^2)P_L + \frac{2}{3}s_W^2P_R]$ & $Y^0_{\mu} \overline{D} d:$  & $-\frac{ig}{\sqrt{2}} \gamma_{\mu} P_L $ \\
\hline
$Z_h^\mu \overline d d:$ & $-\frac{g}{c_W \sqrt{3-4s_W^2}}\gamma^{\mu}[(-\frac{1}{2} + \frac{2}{3}s_W^2)P_L - \frac{1}{3}s_W^2P_R]$ & $X^-_{\mu} \overline{e} N:$  & $ -\frac{ig}{\sqrt{2}} \gamma_{\mu} P_L $ \\
\hline
$Z_h^\mu \overline e e:$ & $-\frac{g}{c_W \sqrt{3-4s_W^2}}\gamma^{\mu}[(\frac{1}{2} - s_W^2)P_L  - s_W^2P_R]$ & $Y^0_{\mu} \overline{\nu} N:$ & $ -\frac{ig}{\sqrt{2}} \gamma_{\mu} P_L $ \\
\hline
$Z_h^\mu \overline{T} T:$ & $-\frac{g}{c_W \sqrt{3-4s_W^2}}\gamma^{\mu}[(-1 + \frac{5}{3}s_W^2)P_L + \frac{2}{3}s_W^2P_R]$ & $X^-_{\mu} \overline{b} T:$   & $ -\frac{ig}{\sqrt{2}} \gamma_{\mu} P_L$ \\
\hline
$Z_h^\mu \overline{D} D:$ & $-\frac{g}{c_W \sqrt{3-4s_W^2}}\gamma^{\mu}[(1 - \frac{4}{3}s_W^2)P_L - \frac{1}{3}s_W^2P_R]$ & $Y^0_{\mu} \overline{t} T:$   & $ -\frac{ig}{\sqrt{2}} \gamma_{\mu} P_L$ \\
\hline
\end{tabular}
\caption{\label{fermiongb}Feynman rules for gauge boson-fermion couplings in the Simplest Little Higgs with T-parity model. $P_L = (1 - \gamma^5)/2$ and $P_R = (1 + \gamma^5)/2$ are projection operators. Couplings for the second generation quarks are identical to those of the first generation. The SM lepton $Z^0$-boson coupling is also provided to clarify our phase choice. Only the leading terms are provided.}
\end{table*}

\begin{table*}[!tbp]
\centering
\vspace{10pt}
\begin{tabular}{|c|c|}
\hline
Particles & Vertices \\
\hline
$W^+W^-h:$ & $\frac{1}{2}g^2v$\\
\hline
$X^+X^-h:$ & $-\frac{1}{2}g^2v$\\
\hline
$X^+X^-hh:$ & $-\frac{1}{2}g^2$\\
\hline
$Y^0\bar{Y}^0h:$ & $ \frac{g^2v}{8}\frac{v^2}{f^2}$\\
\hline
$Y^0\bar{Y}^0hh:$ & $ \frac{g^2}{8}\frac{v^2}{f^2}$\\
\hline
$Z_hZ_hh:$ & $-\frac{g^2v}{2c_W^2}$\\
\hline
$Z_hZ_hhh:$ & $-\frac{g^2}{2c_W^2}$\\
\hline
$Z^0Z_hh:$ & $\frac{1}{2}g^2v\frac{(1-t_W^2)}{c_W\sqrt{3-t_W^2}}$\\
\hline
$Z^0Z_hhh:$ & $\frac{1}{2}g^2\frac{(1-t_W^2)}{c_W\sqrt{3-t_W^2}}$\\
\hline
\end{tabular}
\caption{\label{vvs}Feynman rules for gauge boson-scalar couplings in the Simplest Little Higgs with T-parity model. The SM coupling of the $W^{\pm}$ to the Higgs field $h$ is also provided for reference. Factor of $g_{\mu\nu }$ is implied in each vertex.}
\end{table*}

The fermion sector of the Simplest Little Higgs with T-parity model is very different from the original Simplest Little Higgs. The modifications are especially significant in the Yukawa interactions. Things are also complicated by the fact that, choosing the anomaly-free implementation of fermion assignments, the third generation quarks transform differently from the quarks of the first two generations under the global symmetry of the model. For the purposes of dark matter study we ignore mixings in the leptonic and quark sectors. This leads to a considerable simplification in the Feynman rules, and allows us to concentrate on the most significant ``smoking-gun'' signatures of the model. The more complete set of couplings including $v^2/f^2$ corrections and various mixing can be derived from generic Simplest Little Higgs Lagrangians provided in~\cite{Hanslht}  after imposing the T-parity requirements described in Section 2. In Table~\ref{Hll} we show scalar-fermion couplings that arise from the Yukawa Lagrangians introduced in Section 2.

The fermion couplings to the gauge bosons are determined by the standard fermion kinetic term:
\begin{displaymath}
        \mathcal{L} = \bar \psi i \mathcal{D}_{\mu} \gamma^{\mu} \psi,
	\qquad
	\mathcal{D} = \partial + i g A^a T^a + i g_x Q_x B^x,
\end{displaymath}
with the $Q_x$ charges given in Table~\ref{slhtrpn}. $T^a$ are the generators of the fundamental $\mathbf{3}$ representation of the $SU(3)_W$. In the case of the first and second generation quarks $T^a$ should be replaced by $-T^{a*}$ -- generators of the antifundamental $\mathbf{\bar{3}}$ representation of the $SU(3)_W$. Fermion couplings to the gauge bosons in the SLHT are identical to those of the original Simplest Little Higgs model  and are provided in Table~\ref{fermiongb}. We note, however, that our value for the coupling of $D$ and $S$ quarks to the heavy $Z_h$ is slightly different from the coupling calculated in~\cite{Hanslht}. 

Finally, in Table~\ref{vvs}, we provide the couplings of the new heavy gauge bosons and the scalar Higgs boson. We also note that, because the new electroweak singlet $\eta$ is a real scalar field, it does not couple directly to the gauge bosons, while its quartic couplings are $f^2$ suppressed. Therefore, for the investigation of dark matter phenomenology, effects of the new heavy scalar were completely negligible.%

\vfill
\eject
\newpage
\bibliographystyle{prsty}

\end{document}